\documentclass[11pt,a4paper]{article}

\usepackage{appendix}

\usepackage{graphicx}% Include figure files
\usepackage{epsfig}

\usepackage{hyperref}
\usepackage{graphicx,epsf}
\usepackage{bm}
\usepackage{amsmath}

\newcommand{\vecb}[1]{{\bf #1}}
\def \d {\mathrm{d}}

\begin{document}

\begin{center}
{\large{\textbf{Effect of the electron redistribution on the nonlinear saturation of Alfv\'en eigenmodes and the excitation of zonal flows}}}\\
\vspace{0.2 cm}
{\normalsize {A. Biancalani$^1$, A. Bottino$^1$, P. Lauber$^1$, A. Mishchenko$^2$, and F. Vannini$^1$.\\}}
\vspace{0.2 cm}
\small{$^1$ Max Planck Institute for Plasma Physics, 85748 Garching, Germany\\
$^2$ Max Planck Institute for Plasma Physics, 17491 Greifswald, Germany}\\
\vspace{0.1 cm}
{\footnotesize{contact of main author: \url{http://www2.ipp.mpg.de/~biancala/}}}
%created 2018-03-07\\
%last modified 2019-10-24}
\end{center}

\begin{abstract}
Numerical simulations of Alfv\'en modes (AM) driven by energetic particles are performed with the GK global PIC code ORB5. A reversed shear equilibrium magnetic field is adopted. A simplified configuration with circular flux surfaces and large aspect ratio is considered. The nonlinear saturation of beta-induced Alfv\'en eigenmodes (BAE)  is investigated. The roles of the wave-particle nonlinearity of the different species, i.e. thermal ions, electrons, and energetic ions are described, in particular for their role in the saturation of the BAE and the generation of zonal flows. The nonlinear redistribution of the electron population is found to be important in increasing the BAE saturation level and the zonal flow amplitude.
\end{abstract}

%\vskip 6em

\section{Introduction}
\label{sec:intro}

Tokamak plasmas often present a supra-thermal species, due to external heating or to the product of fusion reactions. This energetic particle (EP) species, can linearly and nonlinearly interact with plasma oscillations via inverse Landau damping.
Alfv\'en modes (AM) are transverse electromagnetic plasma oscillations propagating along the equilibrium magnetic field in magnetized plasmas. In tokamaks, AMs are important as they can effectively tap the free energy of the EP population, nonlinearly modifying the EP distribution~\cite{Kolesnichenko67,Rosenbluth75,Chen16}.
The theoretical prediction of the saturation level of AMs is of crucial importance as we are in the phase of building future fusion devices, where AMs might strongly redistribute the energetic particle (EP) population, affecting, inter alia, the heating efficiency. 

One of the main saturation mechanisms of AMs is the wave-particle nonliner interaction with the EPs. Due to the AM field, the EP trajectory is modified and the EP distribution in phase space evolves, relaxing the drive intensity and leading to the decrease of the AM growth rate. This nonlinear EP dynamics has been studied extensively in the last decades. Due to the focus on the EP dynamics, most of these studies have been carried out with global simulations neglecting the nonlinear kinetic dynamics of both the thermal ions and electrons~\cite{Berk90, Todo93, Briguglio98, Schneller12, Biancalani17, Cole17}, and some of them including the nonlinear kinetic dynamics of the ions~\cite{Zhang13pop, Todo19}. In this paper, we investigate the separate contributions of the wave-particle nonlinearity of the different species, with emphasis on the importance of the nonlinear kinetic dynamics of the electrons. In particular, we show how the saturation level of the AM is affected by the nonlinear radial redistribution of the electrons, which causes a relaxation of the electron Landau damping.

Another possible saturation mechanism of AMs is given by the wave-wave coupling, i.e. the energy exchange among different modes. A particularly interesting energy channel is given by the nonlinear generation of a zonal, i.e. axisymmetric, flows (ZF). This has been investigated numerically~\cite{Todo10, Zhang13pst} and analytically~\cite{Chen12,Qiu-NuFu-16}. A force-driven excitation has been identified numerically~\cite{Todo10} and explained analytically~\cite{Qiu-NuFu-16}. Although the mechanism of the force-driven excitation is mediated by the EPs, and therefore does not require the electrons to be treated kinetically, nevertheless, the correct estimation of the evolution in time of the amplitude of the AM depends on the electrons, as mentioned above. As a consequence, also the nonlinear dynamics of the ZF depends on the inclusion of kinetic electrons.

The numerical investigation described here is carried out with the gyrokinetic particle-in-cell code ORB5. ORB5 was originally written for electrostatic turbulence studies~\cite{Jolliet07}, and recently extended to its electromagnetic multispecies version~\cite{Bottino11,Lanti19}. The global character of ORB5, i.e. the resolution of the full radial extension of the global eigenmodes to scales comparable with the minor radius, in contrast with local ``flux-tube'' models, makes ORB5 appropriate for studying low-n AMs (without pushing towards the local limit of vanishing ratios of the ion Larmor radius to the tokamak minor radius).
The recent inclusion of the pull-back scheme for the mitigation of the cancellation problem greatly improved the efficiency of ORB5~\cite{Mishchenko18}, making these simulations feasable.
 
The paper is structured as follows. In Sec.~\ref{sec:model}, the governing equations of the ORB5 model are given. In Sec.~\ref{sec:equil-profs}, the considered magnetic equilibrium and the plasma profiles are described. The linear dynamics of this configuration is depicted in Sec.~\ref{sec:linear}. In particular, a beta-induced Alfv\'en Eigenmode (BAE)~\cite{Chu92,Heibrink93} is found to be the dominant AM in this configuration. The description of the nonlinear simulation is given in Sec.~\ref{sec:nonlinear-1}. In Sec.~\ref{sec:nonlinear-2}, we investigate the contribution of the different species to the evolution in time of the AM amplitude and ZF amplitude. Finally, the conclusions and outlook are given in Sec.~\ref{sec:conclusions}.

\section{The model}
\label{sec:model}

The gyrokinetic particle-in-cell code ORB5 is based upon a set of model equations derived by a gyrokinetic Lagrangian~\cite{Bottino15JPP,Tronko16,Lanti19}. These equations are the gyrocenter trajectories, and the two equations for the fields.

The gyrocenter trajectories are:
\begin{eqnarray}
\dot{\vecb  R}&=&\frac{1}{m}\left(p_\|-\frac{e}{c}J_0A_\parallel\right)\frac{\vecb{B^*}}{B^*_\parallel} + \frac{c}{e B^*_\parallel} \vecb{b}\times \left[\mu
  \nabla B + e \nabla J_0  \big(\phi -  \frac{p_\|}{mc} A_\| \big) \right] \label{eq:trajectories_a} \\
\dot{p_\|}&=&-\frac{\vecb{B^*}}{B^*_\parallel}\cdot\left[\mu \nabla B + e
  \nabla J_0 \big(\phi -  \frac{p_\|}{mc} A_\| \big) \right] \label{eq:trajectories_b}
\end{eqnarray}
Here, the phase-space coordinates are $\vecb{Z}=(\vecb{R},p_\|,\mu)$, i.e. respectively the gyrocenter position, canonical parallel moment $p_\| = m U + (e/c) J_0 A_\|$ and magnetic momentum $\mu = m v_\perp^2 / (2B)$.
The Jacobian is given by the parallel component of $\vecb{B}^*= \vecb{B} + (c/e) p_\| \vecb{\nabla}\times \vecb{b}$, where $\vecb{B}$ and $\vecb{b}$ are the equilibrium magnetic field and magnetic unit vector.
The time-dependent fields are named $\phi$ and $A_\|$, and they are respectively the perturbed scalar potential and the parallel component of the perturbed vector potential.
In our notation, on the other hand, $\vecb{A}$ is the equilibrium vector potential.  The summation is over all species present in the plasma, and the gyroaverage operator is labeled here by $J_0$ (with $J_0=1$ for electrons). The gyroaverage operator reduces to the Bessel function if we transform into Fourier space.
All ion species are treated with a gyrokinetic model, whereas electrons are treated with a drift-kinetic model. In other words, we take into account finite-Larmor radius of ions, and we neglect them for electrons. Finite orbit widths are taken into account for all species.
In this paper, we will describe simulations where some species are allowed to redistribute in phase space, i.e. where the corresponding markers are pushed along the perturbed trajectories, Eqs.~\ref{eq:trajectories_a} and \ref{eq:trajectories_b}, and other species are not allowed to redistribute in phase space, i.e. where the corresponding markers are pushed along unperturbed trajectories, i.e. Eqs.~\ref{eq:trajectories_a} and \ref{eq:trajectories_b} without the terms proportional to $\phi$ and $A_\|$.

The GK Poisson equation is:
\begin{equation}
 - \vecb{\nabla} \cdot \frac{mc^2\int \d W f_M}{B^2} \nabla_\perp \phi=\Sigma_{\rm{sp}} \int \d W  e J_0 f  \label{eq:Poisson}
\end{equation}
Here $f$ and $f_M$ are the total and equilibrium (i.e. independent of time) distribution functions, the integrals are over the phase space volume, with $\d V$ being the real-space infinitesimal and $\d W = (2\pi/m^2) B_\|^* \d p_\| \d \mu$ the velocity-space infinitesimal.

Finally, the Amp\`ere equation is:
\begin{eqnarray}
%  &&\int \d W \left( \Sigma_{\rm{sp}}
% \frac{ep_\|}{mc}J_0f-\Sigma_{\rm{sp}}\frac{e^2}{mc^2}A_\parallel f_{M}
% -\Sigma_{\rm{sp\ne
%     e}}\left(\frac{k_BT}{4 B^2}f_M\right)\nabla_\perp^2A_\parallel\right. \nonumber \\ 
% &&
%  -{\color{blue}\Sigma_{\rm{sp\ne       e}}\frac{1}{B^*_\parallel}\nabla_\perp^2\left(B^*_\parallel\frac{k_BT}{ 4 B^2}f_MA_\parallel\right)}
%     +\frac{1}{4\pi}\nabla_\perp ^2 A_\parallel \Big)=0 \label{eq:Amp\`ere}
\Sigma_{\rm{sp}} \int \d W \Big( \frac{ep_\|}{mc} f-\frac{e^2}{mc^2}A_\parallel f_{M}
%-\Sigma_{\rm{sp\ne e}}\Big(\frac{k_BT}{4 B^2}f_M\Big)\nabla_\perp^2A_\parallel 
 %-{\color{blue}\Sigma_{\rm{sp\ne       e}}\frac{1}{B^*_\parallel}\nabla_\perp^2\Big( B^*_\parallel\frac{k_BT}{ 4 B^2}f_MA_\parallel \Big)}
 \Big)  +  \frac{1}{4\pi}\nabla_\perp ^2 A_\parallel =0 \label{eq:Ampere}
\end{eqnarray}
Here the form with $J_0=1$ is given for simplicity, also in the view of the comparison with MHD codes. For more complicated models, see Ref.~\cite{Tronko16,Lanti19}. The pull-back scheme is used, to mitigate the cancellation problem~\cite{Mishchenko18}.

\section{Magnetic equilibrium and plasma profiles}
\label{sec:equil-profs}

The tokamak geometry and magnetic field are taken consistently with Ref.~\cite{Biancalani16PoP}, for the case with reversed shear. The major radius is $R_0 = 1.0$ m, the minor radius is $a=0.1$ m, and the toroidal magnetic field at the axis is $B_0 = 3.0$ T. Circular concentric flux surfaces are considered. The safety factor has a value of 1.85 at the axis, it decreases from $\rho$=0 to $\rho$=0.5, where the minimum value is located ($q(\rho=0.5)$=1.78), and then it raises to the edge, where it reaches the maximum value ($q(\rho=1)$=2.6). Here $\rho$ is a normalized radial coordinate defined as $\rho=r/a$.

\begin{figure}[b!]
\begin{center}
\includegraphics[width=0.45\textwidth]{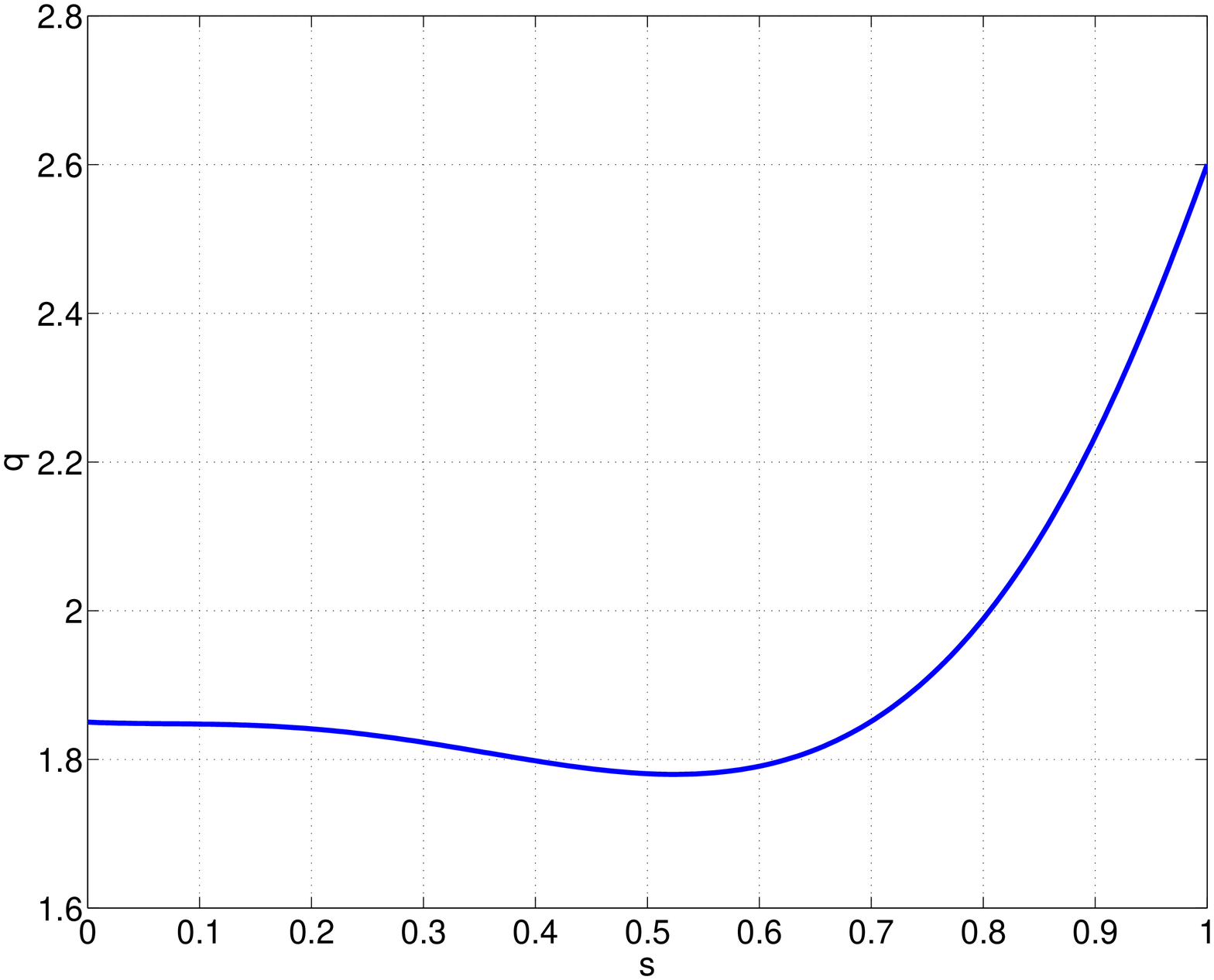}
%\includegraphics[width=0.45\textwidth]{equil-q_s-zoom.eps}
%\includegraphics[width=0.49\textwidth]{q1_4-gamma_e-k2.eps}
%\vskip -1em
\caption{\label{fig:q_s} Safety factor profile.}
\end{center}
\end{figure}

We choose a reference radial position of $\rho=\rho_r=0.5$, corresponding to $s=0.525$, where the flux radial coordinate $s$ is defined as $s=\sqrt{\psi_{pol}/\psi_{pol}(edge)}$, $\psi_{pol}$ being the poloidal magnetic flux. The ion and electron temperatures are taken equal everywhere, $T_e(\rho)=T_i(\rho)$. Here, differently from Ref.~\cite{Biancalani16PoP}, a value of $T_e(\rho=\rho_r)$ corresponding
%to $Lx=2/\rho^* = 350$, i.e. 
$\rho^* = \rho_s/a = 0.00571$, is chosen (with $\rho_s = \sqrt{T_e/m_i}/\Omega_i$ being the sound Larmor radius).
The electron thermal to magnetic pressure ratio of $\beta_e = 8\pi \langle n_e \rangle T_e(\rho_r)/B_0^2 = 5e-4$.
%(therefore, in ORB5, we set \verb|beta = 2.5e-4|). Here $\langle n_e \rangle = $ \verb|nbar|$_e * n_e(0)$, with \verb|nbar|$_e$ = 0.971 given in ORB5's standard output, and depends on the choice of $\kappa_n$ (see Eq.~\ref{eq:n_EP}) relative to the electron density profile.
% 
% \begin{figure}[t!]
% \begin{center}
% \includegraphics[width=0.44\textwidth]{equil-s_rho.eps}
% \includegraphics[width=0.45\textwidth]{equil-s_rho-zoom.eps}
% %\includegraphics[width=0.49\textwidth]{q1_4-gamma_e-k2.eps}
% %\vskip -1em
% \caption{\label{fig:s_rho} Mapping of the two radial coordinates $s$ and $\rho$ (left) and zoom near the minimum of the safety factor (right).}
% \end{center} 
% \end{figure}

We consider a deuterium plasma. This means that the chosen value of $\rho^*$ corresponds to a temperature at the reference position of $T_e(\rho_r)=14.05$ keV.
This also corresponds to an electron density on axis of $n_e(0)=0.818e18$ m$^{-3}$ and at the reference radius of $n_e(\rho_r)=0.775e18$ m$^{-3}$.
Some characteristic frequencies and velocities can be calculated. The ion cyclotron frequency is $\Omega_i = 1.44 e8$ rad/s. The Alfv\'en velocity on axis is $v_A(0)=5.12 e7$ m/s.
The sound velocity at the reference radius is $c_s(\rho_r) =\sqrt{T_e(\rho_r)/m_i}= 8.21e5$ m/s. 
The Alfv\'en frequency on axis is $\omega_A = v_A(0) / q(0)R_0 = 2.76e6$ rad/s $= 1.92$e-2 $\Omega_i$. The sound frequency on axis is $\omega_s(0) = \sqrt{2} c_s(0) / R_0 = 1.16e5$ rad/s. 

\begin{figure}[t!]
\begin{center}
\includegraphics[width=0.45\textwidth]{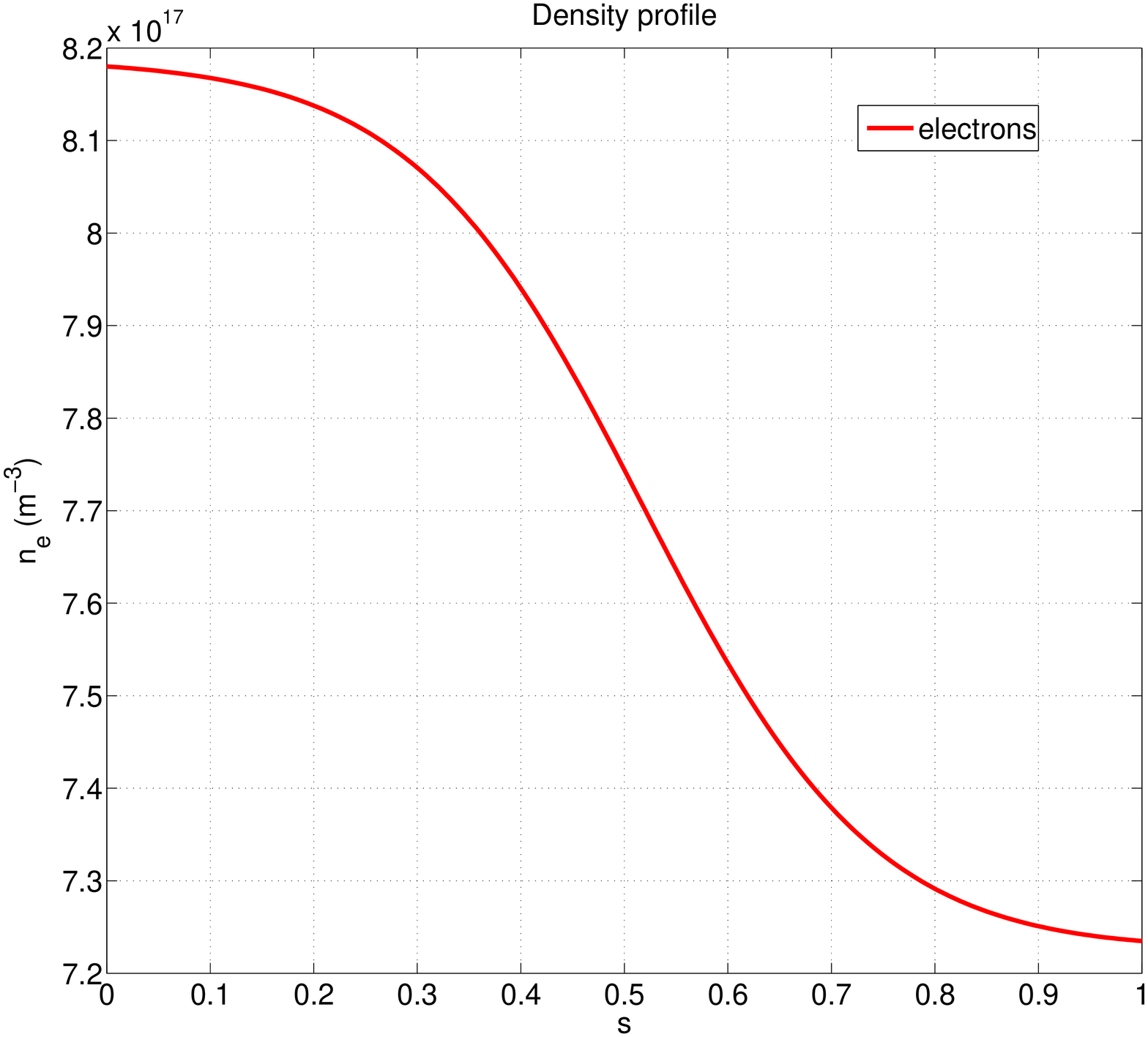}
\includegraphics[width=0.45\textwidth]{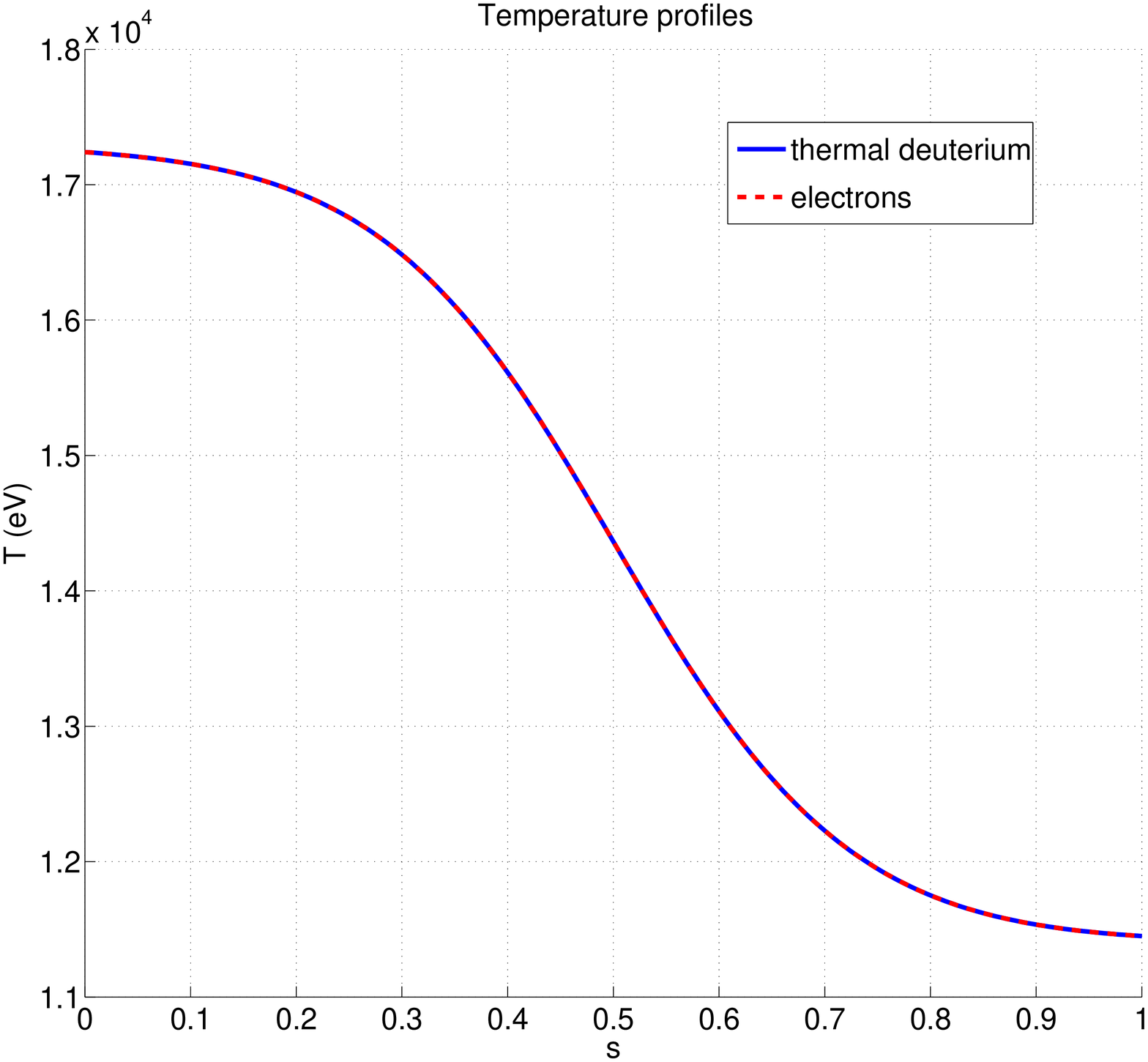}
%\includegraphics[width=0.49\textwidth]{q1_4-gamma_e-k2.eps}
%\vskip -1em
\caption{\label{fig:profs_s} Density (left) and temperature (right) profiles, vs $s$ radial coordinate.}
\end{center} 
\end{figure}

The same analytical function is used for all the profiles of the equilibrium density and temperature, for the three species of interest (thermal deuterium, labelled here as ``d'', thermal electrons, labelled here as ``e'', and hot deuterium, labelled here as ``EP'').
%The profile type in ORB5 is chosen as \verb|nsel_profile=2|, therefore the $\rho$ coordinate is used.
For the EP density, for example, the function is written as:
\begin{equation}\label{eq:n_EP}
n_{EP}(\rho)/n_{EP}(\rho_r) = \exp [-\Delta \, \kappa_n \tanh ((\rho-\rho_r)/\Delta)] 
\end{equation}
The value of $\Delta$ is the same for all species, for both density and temperature: $\Delta = 0.208$.  Deuterium and electrons have $\kappa_n=0.3$ and $\kappa_T=1.0$, and the EP have  $\kappa_n=10.0$ and $\kappa_T=0.0$. Different values of the EP temperature are considered. The value of $\rho_r$ is the reference radial position defined above.

The distribution function of the EP population is Maxwellian. Different values of the EP averaged concentration $\langle n_{EP} \rangle /n_e$ are considered.
Once the EP population is loaded into ORB5, we choose to satisfy the quasineutrality for the considered simulations. This means that ORB5 automatically re-calculates the electron density profile, in order to have $n_{EP} (\rho) +n_i = n_e (\rho)$ for all values of $\rho$.

In the nonlinear simulations shown in this paper, a Fourier filter in mode numbers is applied, keeping  $0\le n \le 9$ and
%$-32 \le m \le 32$.
$|m- n q(s) |\le 5$.
Regarding the radial direction, unicity boundary conditions are imposed to the potentials at the axis and Dirichlet at the external boundary. The value of the electron mass is chosen 200 times lighter than thermal ions. This value is found to be at convergence (see Appendix~\ref{app:mass}).

\section{Linear dynamics}
\label{sec:linear}

\begin{figure}[h!]
\begin{center}
\includegraphics[width=0.54\textwidth]{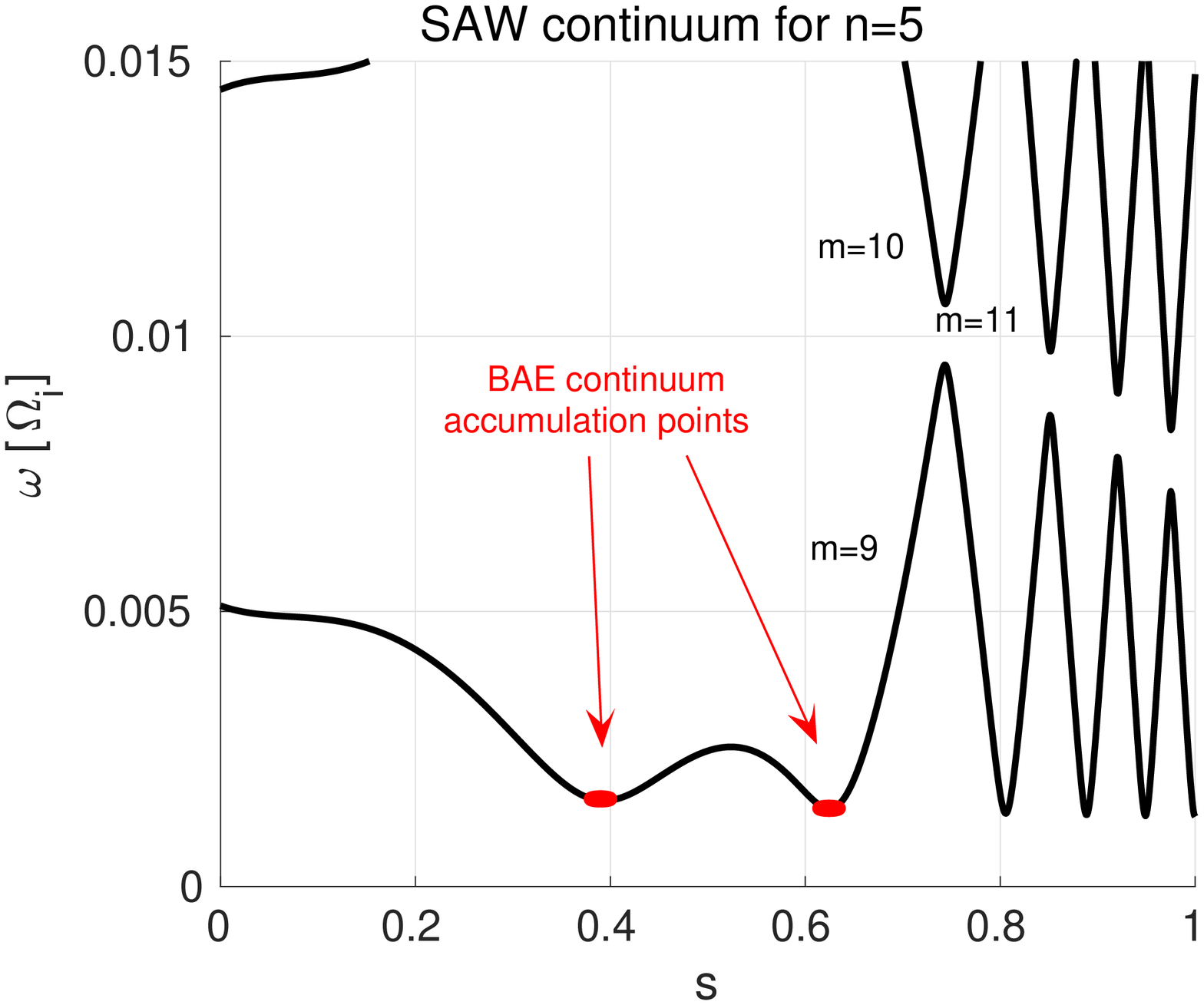}
\includegraphics[width=0.44\textwidth]{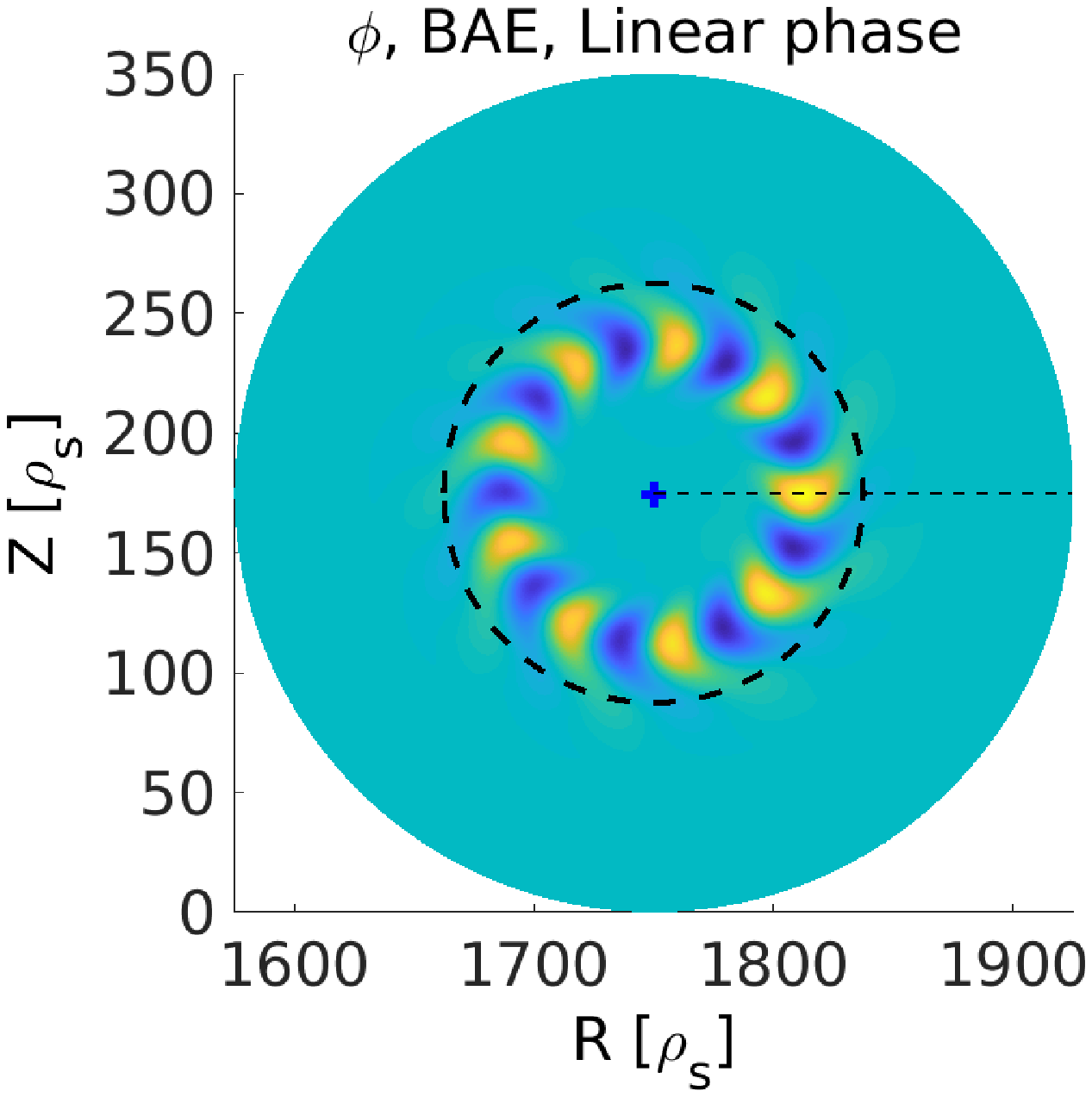}
%\includegraphics[width=0.49\textwidth]{q1_4-gamma_e-k2.eps}
%\vskip -1em
\caption{\label{fig:SAW-lin} On the left, SAW continuous spectrum for for n=5, depicting the position of the two BAE accumulation points. On the right, poloidal structure of mode linearly excited with an EP population with $T_{EP}/T_e = 10$. The black circular dashed line delimits the s=0.5 flux surface.}
\end{center} 
\end{figure}

In this section, we investigate the linear dynamics of AM in the equilibrium described in the previous section.
%, and the EP with $k_n=10.0$.
In Fig.~\ref{fig:SAW-lin}-left, we show the continuous spectrum for the modes with n=5, which are the toroidal mode numbers observed to be dominant. The formula is taken from Ref.~\cite{Biancalani16PoP}, where an approximation of small $\epsilon$ is considered, slightly underestimating the width of the TAE gaps. This should be valid here, as $\epsilon$ is small. The BAE continuum accumulation point (CAP) is calculated with the GAM frequency, due to the BAE-GAM degeneracy~\cite{Zonca08}.

Linear numerical simulations are also performed with ORB5 studying the different toroidal mode numbers. The EP population is initialized with $T_{EP}/T_e = 10$ at mid-radius. The dominant mode is found to be the one with n=5, m=9. The frequency and growth rate measured in a simulation with $\langle n_{EP} \rangle/\langle n_e \rangle = 0.01$ are:
\begin{eqnarray}
 \omega_{lin} &=& 2.4\cdot 10^{-3} \; \Omega_i\\
 \gamma_{lin} &=& 0.68\cdot 10^{-3} \; \Omega_i
\end{eqnarray}
Due to the spatial structure and to the frequency in relation to the SAW continuous spectrum, this mode is identified as the BAE at the inner continuum accumulation point (see Fig.~\ref{fig:SAW-lin}-right). The mode is observed to rotate in the ion-diamagnetic direction. Note that the frequency is observed slightly above the BAE-CAP for finite EP concentrations, analogously with what described in Ref.~\cite{Lauber13}, and goes to the BAE-CAP for vanishing EP concentrations.
% 
% 
% \begin{figure}[h!]
% \begin{center}
% \includegraphics[width=0.48\textwidth]{SAW-lin-gamma_TH-nH10.eps}
% \vskip -1em
% \caption{\label{fig:SAW-lin-scan-gamma} Dependence of the AM growth rate on the EP temperature, for an EP population with $\langle n_{EP}\rangle /n_e = 0.01$. No krook operator is applied here.}
% \end{center} 
% \end{figure}

\newpage
\section{Description of the fully nonlinear simulation}
\label{sec:nonlinear-1}

In this section, we describe the nonlinear dynamics of the Alfv\'en modes in an equilibrium described in Section~\ref{sec:equil-profs}. The analysis of the linear dynamics has been described in Sec.~\ref{sec:linear}, showing that a BAE with n=5, m=9, is the most unstable mode. Here, we allow all modes with toroidal mode number in the interval $[0,9]$ to develop. For completeness, we report here that in the case when only the n=5 mode is allowed to evolve, the nonlinear phase is found to be sensibly different. The simulation is fully nonlinear, meaning that the markers for all species are pushed along perturbed trajectories. No krook operator is applied here, meaning that there are no sources or sinks of energy due to the forcing of the profiles. A perturbation with n=5, m=9 is initialized at t=0. Unicity boundary conditions are applied at the axis, i.e. s=0, and Dirichlet at the edge, i.e. s=1.

\begin{figure}[b!]
\begin{center}
\includegraphics[width=0.48\textwidth]{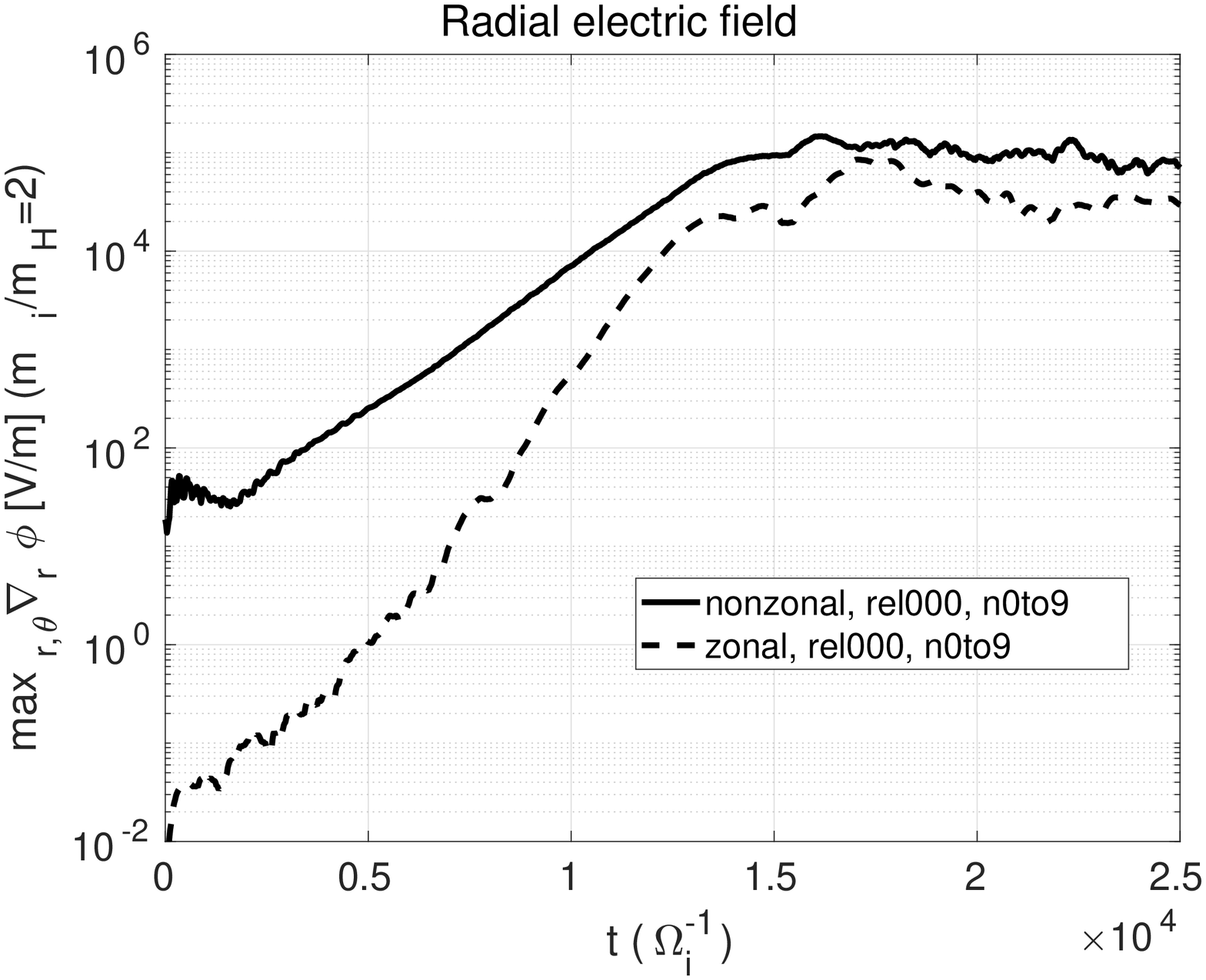}
\includegraphics[width=0.48\textwidth]{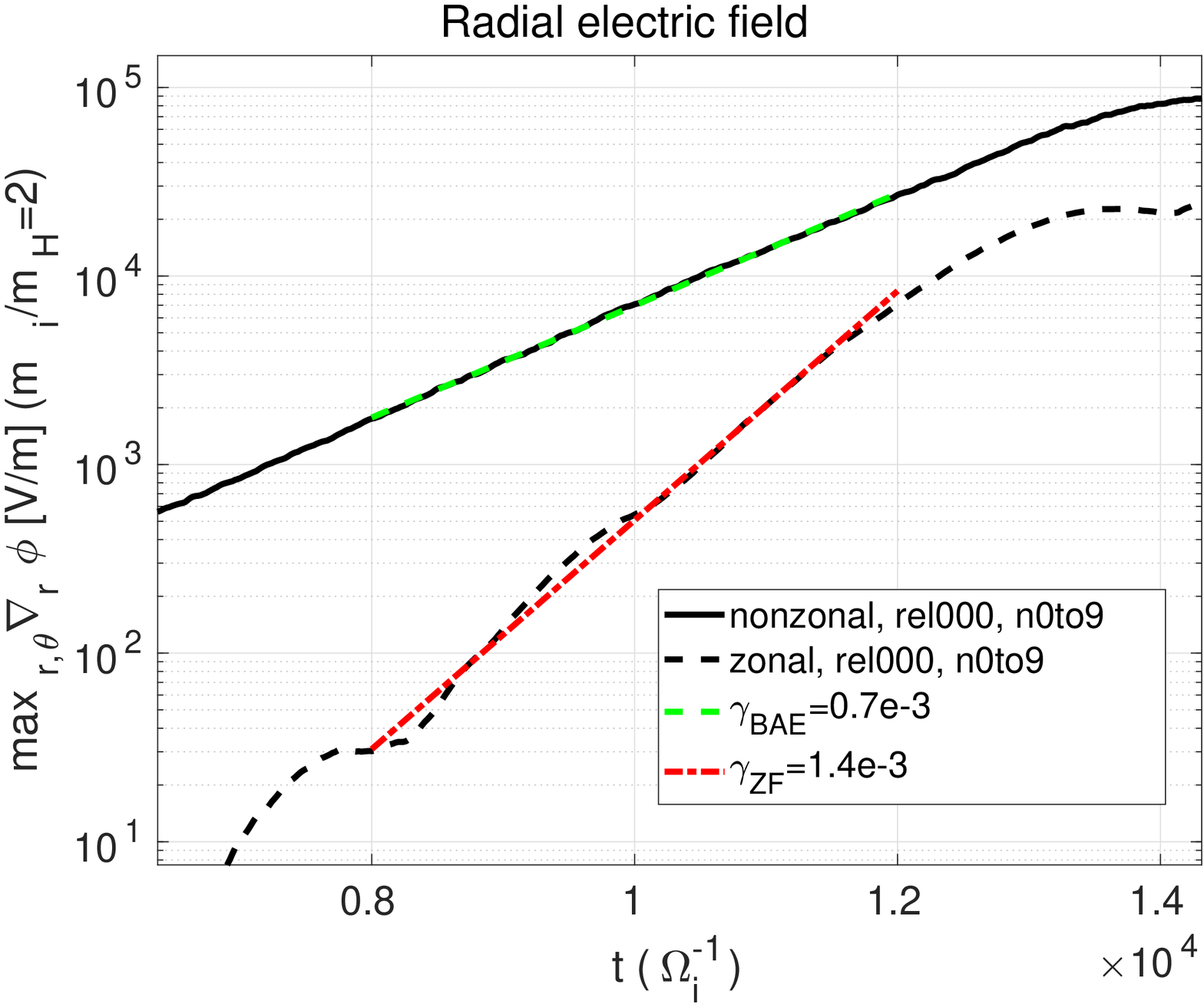}
\vskip -1em
\caption{\label{fig:NL-FULL} On the left, evolution of the nonzonal (continuous) and zonal (dashed) components of the radial electric field. On the right, zoom in the early nonlinear phase.}
\end{center} 
\end{figure}

The zonal component of the radial electric field can be measured as the flux surface average of the total electric field at each radial position. The nonzonal component is defined as the difference of the total radial electric field and the zonal component, at each position in the poloidal plane ($s$,$\theta$). The maximum values of these components are measured in the poloidal plane, at one given toroidal angle ($\varphi=0$), and at each time step, and their evolution in time is shown in Fig.~\ref{fig:NL-FULL}-left. The nonzonal component describes the BAE field, whereas the zonal component describes the zonal flow.
We note that there are 5 phases: 1) a first transient phase, where the BAE mode is forming (0$<$t$<$2000, in $\Omega_i^{-1}$ units); 2) a linear phase, where the BAE grows linearly but the ZF is still at noise levels (2000$<$t$<$7000); 3) an early-nonlinear phase, where the BAE continues growing exponentially, but the ZF starts growing in amplitude with higher growth rate, meaning that a nonlinear coupling is already occurring (7000$<$t$<$13000); 4) an early saturation phase, where the BAE instantaneous growth rate starts decreasing, with respect to the linear value  (13000$<$t$<$16000); 5) a deep saturated phase, after the BAE field has reached the maximum level of $\delta \tilde{E}_r \simeq 1.0e5 $ V/m, and starts decreasing in amplitude (t$>$16000).
In this paper, we focus on the dynamics of the BAE and ZF in the early nonlinear phase and early saturation phase (7000$<$t$<$16000).

The investigation of the early nonlinear phase shows that the ZF starts growing with a growth rate of $\gamma_{ZF}=1.4e-3 \Omega_i$, i.e. double of the BAE linear growth rate. This is a signature of the mechanism named force-driven excitation, observed numerically with MEGA~\cite{Todo10} and GTC~\cite{Zhang13pst} and explained analytically in Ref.~\cite{Qiu-NuFu-16}.

\begin{figure}[t!]
\begin{center}
\includegraphics[width=0.45\textwidth]{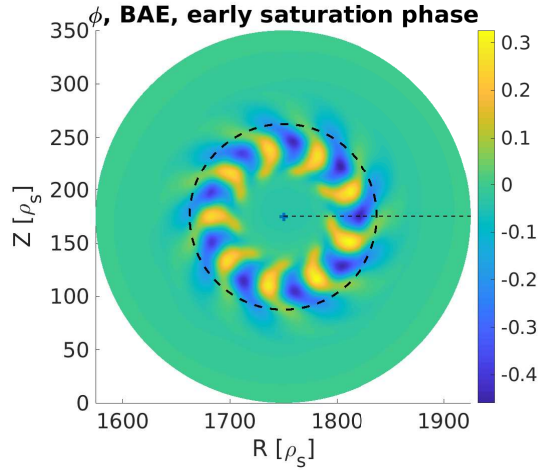}
\vskip -1em
\caption{\label{fig:NL-FULL-struct} BAE spatial structure at t=15000.}
\end{center} 
\end{figure}

The spatial structure of the BAE is observed to change during the early saturation phase. In particular, the radial width becomes larger and the ``arms'' of the ``boomerang'' extend towards the axis and the second BAE CAP (see Fig.~\ref{fig:NL-FULL-struct}). This is a signature of a weaker interaction of the eigenmode with the continuum occurring during the early nonlinear saturation phase, with respect to the early nonlinear phase, consistently with Ref.~\cite{Biancalani17}. Incidentally, we note that the main dynamics of the BAE and of the ZF is not found to sensibly change if the simulations are run with a flat temperature profile, meaning that the global properties of the temperature profiles are not substantial for the analysis of the modes considered here.

\section{Contribution of the different species to the nonlinear dynamics}
\label{sec:nonlinear-2}

During the linear excitation phase, Alfv\'en modes tap the free energy stored in the EP profile. This role of the EP population is well known and it has been studied extensively in literature, as mentioned in the Introduction.
In general, the two ways energy can be exchanged between waves and particles of a given species $s$, are described via nonuniformities in velocity or real space (see, for example, Ref.~\cite{Betti92}). The corresponding contributions to the growth rate have the following dependencies on the distribution function $F_s$ of the species:
\begin{eqnarray}
\gamma_{\epsilon} & \propto & \omega \frac{\partial F_s}{\partial \epsilon}\label{eq:gammaepsilon}\\
\gamma_{\psi} & \propto & - \frac{n}{q_s} \frac{\partial F_s}{\partial \psi}\label{eq:gammapsi}
\end{eqnarray}
where $\epsilon$ is the particle energy, $\psi$ is the flux coordinate, and $q_s$ is the charge of the species $s$. In the case of energetic ions, Eq.~\ref{eq:gammapsi} shows that a negative radial gradient of the density yields a drive. 

\begin{figure}[t!]
\begin{center}
\includegraphics[width=0.48\textwidth]{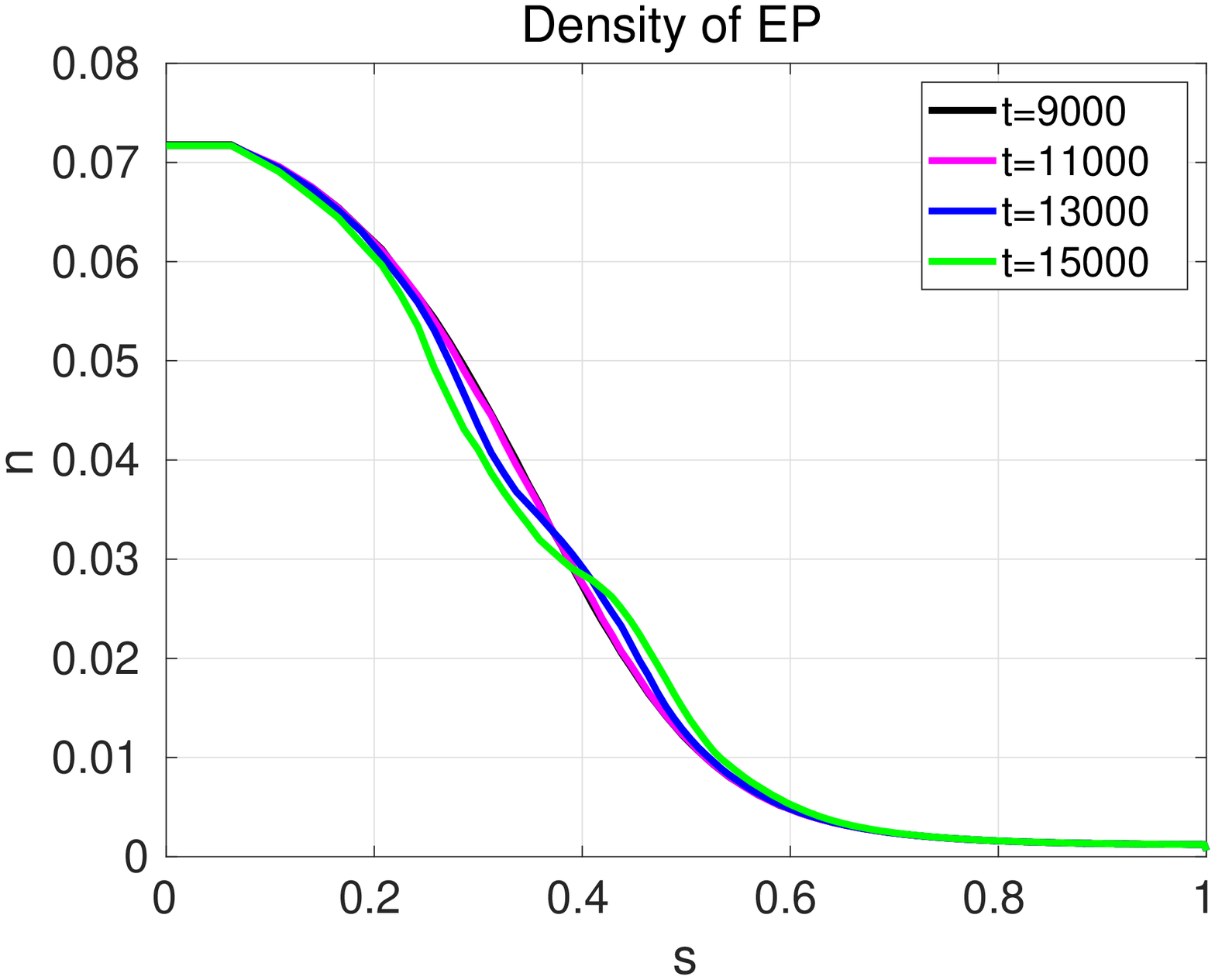}
\includegraphics[width=0.48\textwidth]{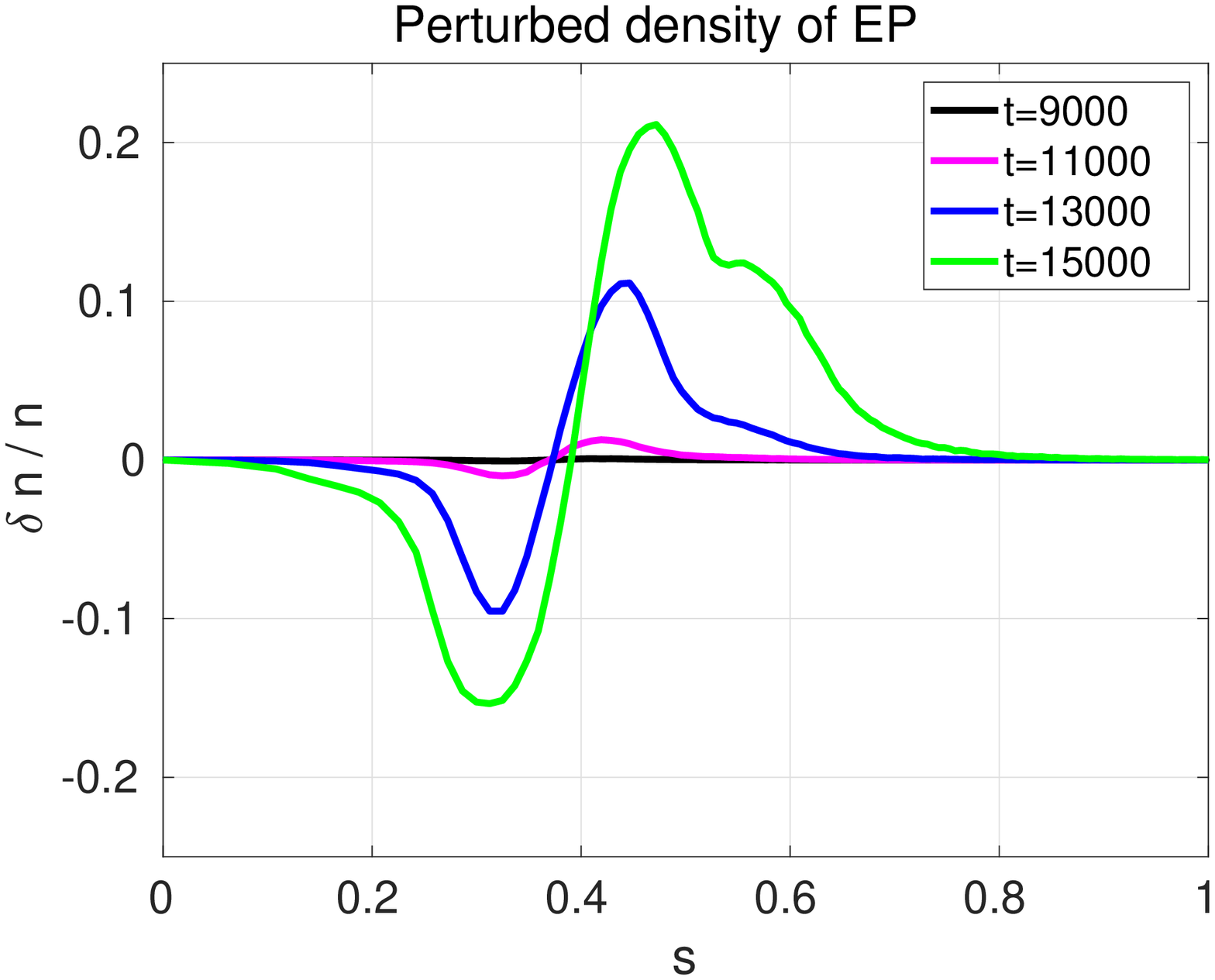}
\vskip -1em
\caption{\label{fig:NL-FULL-EP} On the left, EP profiles depicted at 4 selected times in the early nonlinear phase and early saturation phase. On the right, the corresponding perturbed relative EP density.}
\end{center} 
\end{figure}

During the early saturation phase, the EP population is redistributed radially in the outward direction. 
This phenomenon can be observed also for the BAE observed in the simulations considered in this paper. In Fig.~\ref{fig:NL-FULL-EP}, the outward redistribution of the EP population is shown, around the peak position of the BAE, $s=0.38$. In Fig.~\ref{fig:NL-FULL-EP}-right, note that the characteristic ``hole'' structure forms near the BAE peak, on the side where the EP profile is depleted, whereas the ``clump'' structure forms near the BAE peak, on the side where the EP density increases.

\begin{figure}[b!]
\begin{center}
\includegraphics[width=0.48\textwidth]{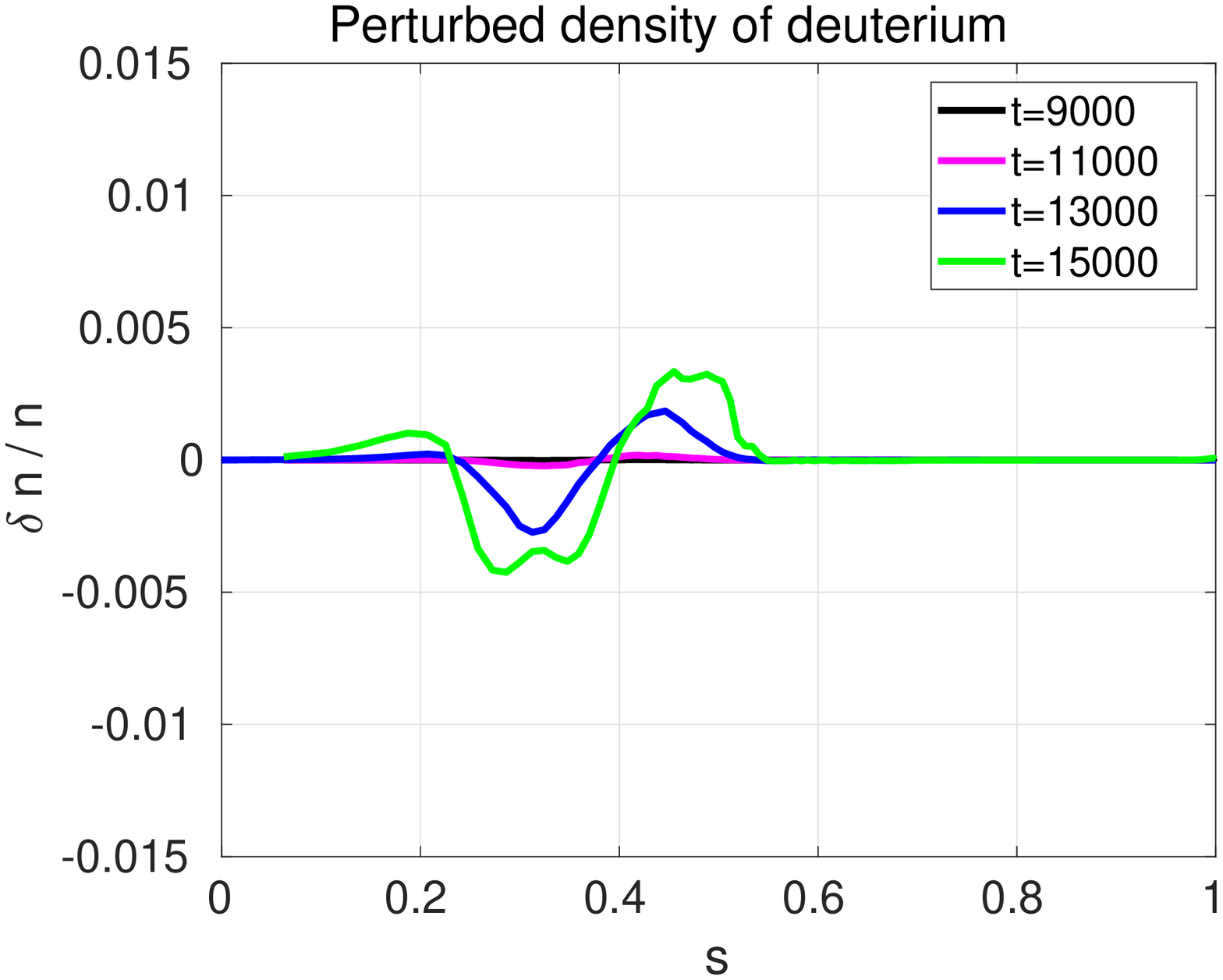}
\includegraphics[width=0.48\textwidth]{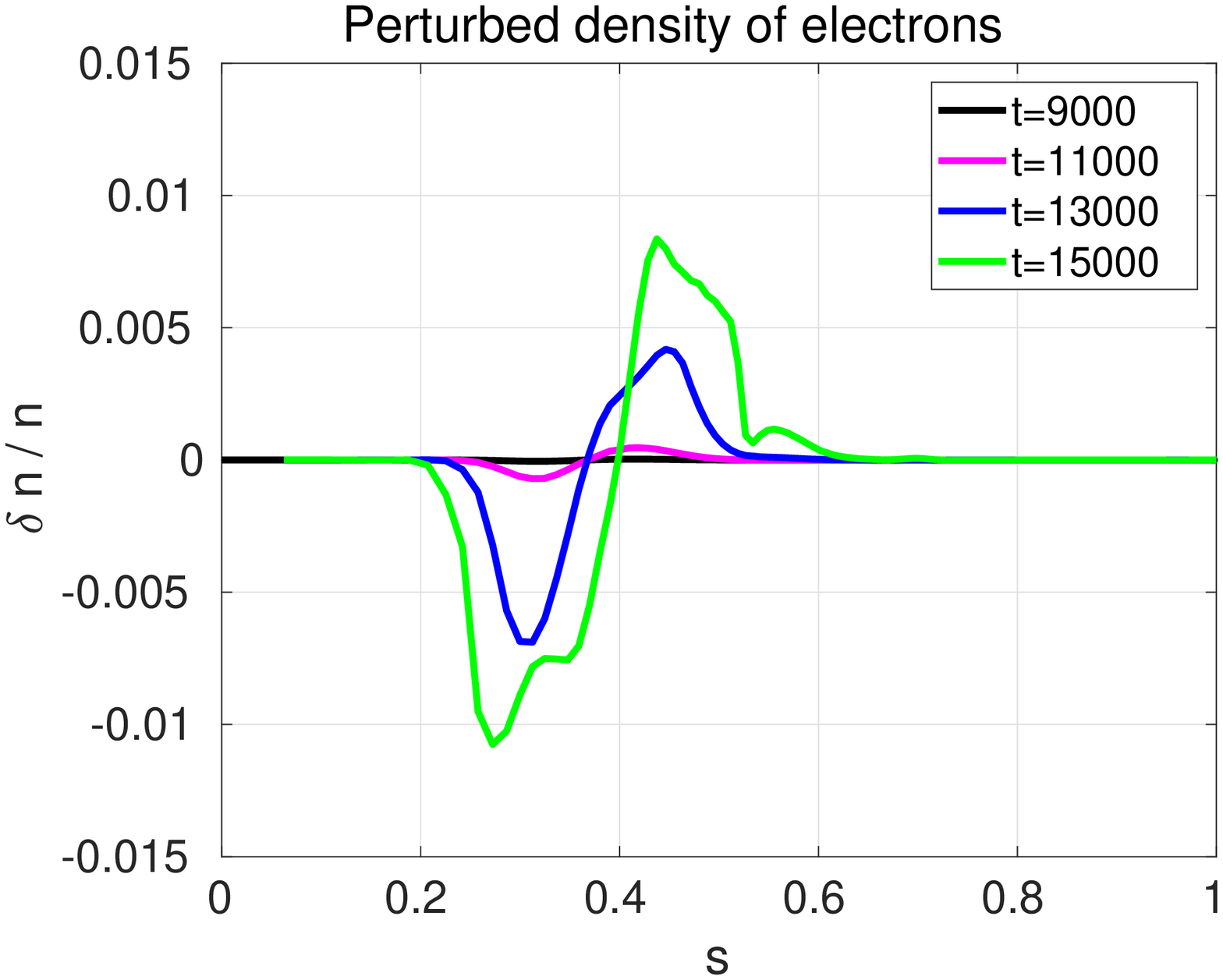}
%\includegraphics[width=0.49\textwidth]{q1_4-gamma_e-k2.eps}
%\vskip -1em
\caption{\label{fig:NL-FULL-deu-ele} Perturbed relative density of the thermal ions (left) and electrons (right).}
\end{center} 
\end{figure}

On the other hand, the role of the redistribution of the thermal species, in the saturation of Alfv\'en modes, has been less investigated in literature. As an example, we now repeat the same study of the perturbed radial density profile done for the EP, for the thermal ions (in our case, deuterium) and electrons. The result is shown in Fig.~\ref{fig:NL-FULL-deu-ele}. Like for the EP, two clear hole and clump structures are observed for the ion and electron species, as well. 
The position of the hole and the clump are the same as for the EP population, denoting the resonance radial location and BAE radial width. Note that the radial extent of the perturbed densities is the same for the ions and electrons, ensuring quasineutrality. 

The thermal ions relative perturbed density is observed to be sensibly smaller than for the electrons, and its effect is also less important, as it is discussed later. On the other hand, the electron perturbed density is not negligible and a strong effect of the electron kinetic dynamics on the nonlinear evolution of the mode is observed in the simulations. Note that the hole and clump observed in the electron perturbed density, corresponding to a flattening of the electron profile, reflect a decreased damping, instead of a decreased drive, which was the case of the EP. This is due to the different charge of the electrons, with respect to the EP, i.e. the energetic deuterium in the case considered here (see Eq.~\ref{eq:gammapsi}). Note also that, in general, both terms described by Eqs.~\ref{eq:gammaepsilon} and \ref{eq:gammapsi} are important and each of them can be the dominant one, in different regimes. In Fig.~\ref{fig:NL-FULL-deu-ele}, the term described in Eq.~\ref{eq:gammapsi} is shown as an example.

\begin{figure}[t!]
\begin{center}
\includegraphics[width=0.47\textwidth]{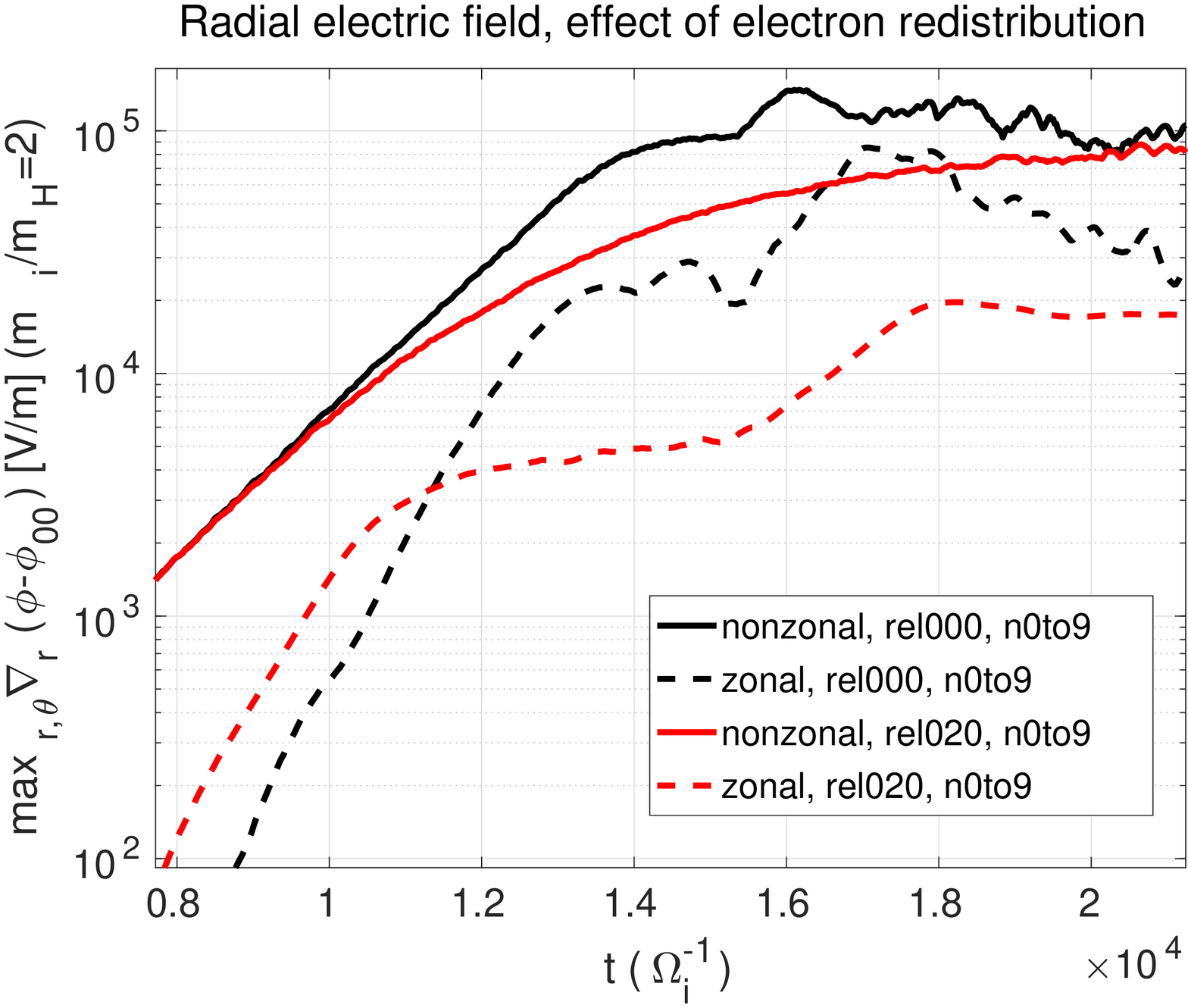}
\includegraphics[width=0.49\textwidth]{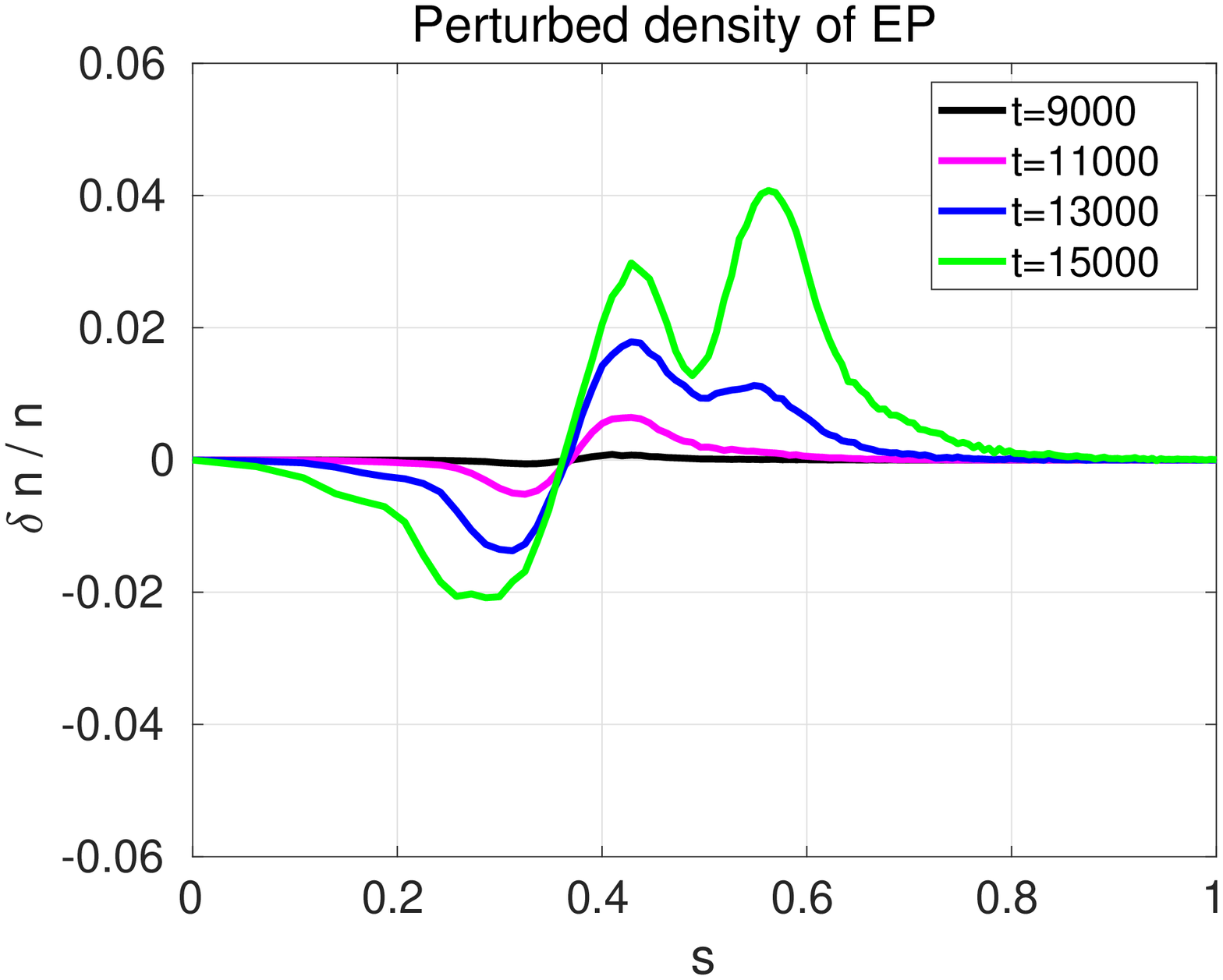}
%\includegraphics[width=0.49\textwidth]{q1_4-gamma_e-k2.eps}
%\vskip -1em
\caption{\label{fig:NL-COMP-1} On the left: evolution in time of the radial electric field nonzonal and zonal components for the fully nonlinear simulation (black lines, also depicted in Fig.~\ref{fig:NL-FULL}) and for a simulation where the electrons are not allowed to redistribute (red lines). On the right: EP perturbed density for the simulation where the electrons are not allowed to redistribute.}
\end{center} 
\end{figure}

In order to check the relevance of the electron redistribution in the nonlinear dynamics of the BAE, we perform a simulation where the electrons are pushed along the unperturbed trajectories. This means that, although the thermal and energetic ions are still allowed to redistribute, the electrons are forced to remain in the initial profile. The result can be observed in Fig.~\ref{fig:NL-COMP-1}-left. We observe that, in the simulation where the electrons are not allowed to redistribute, the early saturation phase starts earlier, namely at t=9000 instead of t=13000, and the BAE amplitude grows more slowly. The cause is a stronger damping due to the electrons. The effect is that, at a given time, the EP redistribution is much lower than in the fully nonlinear case. This is shown in Fig.~\ref{fig:NL-COMP-1}-right.

A simulation with electrons and EPs pushed along perturbed trajectories and thermal ions pushed along unperturbed trajectories has also been performed, and no sensible difference has been found with respect to the fully nonlinear case. This proves that the nonlinear kinetic dynamics of the ions is not relevant in this regime.

\begin{figure}[t!]
\begin{center}
\includegraphics[width=0.47\textwidth]{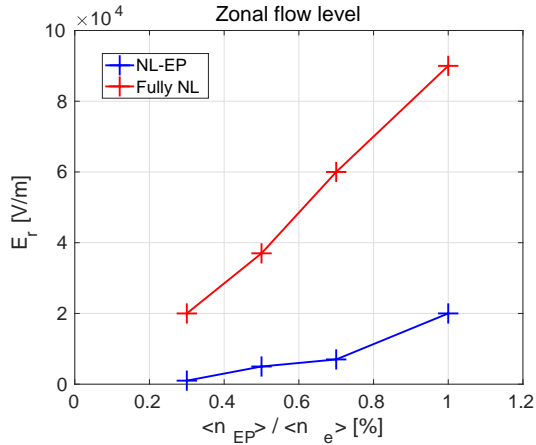}
%\includegraphics[width=0.49\textwidth]{q1_4-gamma_e-k2.eps}
%\vskip -1em
\caption{\label{fig:zonalEr_nEP} Level of the zonal electric field measured around the time of the BAE saturation, for simulations where only EP are allowed to redistribute nonlinearly (blue line), and for fully nonlinear simulations, i.e. where all species are allowed to redistribute (red line).}
\end{center} 
\end{figure}

Finally, we want to investigate how the electron redistribution affects the excitation of zonal structures. In Fig.~\ref{fig:NL-COMP-1}, the evolution in time of the zonal radial electric field is shown for the fully nonlinear case (labelled ``rel000''), and for the case where the electrons are not allowed to redistribute (labelled ``rel020''). The zonal radial electric field is found to reach sensibly higher values in the fully nonlinear case. The reason is the zonal flow is found to follow the analytical prediction of the force-driven excitation, with double growth rate with respect to the BAE, only during the early nonlinear phase. In the simulation where the electrons are not allowed to redistribute, the early saturation phase starts earlier, and the zonal flow growth rate drops to small values earlier, leaving the ZF at lower levels. This is not a threshold-effect, but it increases linearly with the EP drive (see Fig.~\ref{fig:zonalEr_nEP}).

We also note that the growth rate of the ZF during the early nonlinear phase is twice that of the BAE, even in simulations where either one or both the thermal species are not allowed to redistribute. This confirms that, in the early nonlinear phase, the force-driven excitation is mediated dominantly by the nonlinearity associated with the EPs, as predicted in Ref.~\cite{Qiu-NuFu-16}. The main result described in this paper is that the situation changes in the early saturation phase, where the nonlinearity associated with the thermal species, and especially with the electrons, becomes important.

\section{Conclusions}
\label{sec:conclusions}

Alfv\'en modes (AM) are present in tokamak plasmas heated by energetic particles (EP) and they are expected in future fusion devices. Due to their strong nonlinear interaction with the EP population, they are considered with great attention in the theoretical models which aim at being comprehensive. In particular, understanding the saturation level of Alfv\'en modes (AM) in present tokamak plasmas, and predicting it in future fusion devices, is one of the main tasks of the present theoretical community.

The main focus of the analytical and numerical investigation of AMs of the past decades has been the nonlinear wave-particle interaction of AMs and EPs. This has been considered as the main saturation mechanism of AMs. Here, we extend this focus to the wave-particle interaction of AMs with the thermal species, which has been neglected in the past. The profiles of thermal ions and electrons are allowed to redistribute.

For simplicity, a magnetic configuration with circular flux surfaces and large aspect ratio is considered. The beta-induced Alfv\'en Eigenmode (BAE) is observed to be linearly unstable in this configuration, and its linear and nonlinear evolution is studied in simulations where all modes from toroidal mode number n=0 to n=9 are included. The inclusion of the axisymmetric mode n=0 allows us to investigate as well the excitation of zonal flows. The evolution of the BAE is found to be different in simulations where only one toroidal mode is allowed to evolve.

This numerical investigation is performed with the global gyrokinetic particle-in-cell code ORB5. ORB5 was written for turbulence studies, and therefore a sensible effort was put in building a model containing a selfconsistently nonlinear set of equations. The extension of ORB5 to its present electromagnetic multi-species version, allows to investigate global electromagnetic modes like low-n AMs. The gyrokinetic framework, on the other hand, allows to have all the main kinetic effects, for all species, automatically included in the AM dynamics.

The result of this investigation, is that the nonlinear kinetic dynamics of the thermal species, and in particular of the electrons, cannot be neglected if one wants to correctly estimate the saturation level of the BAE and of the force-driven ZF. Although the ions give a small contribution, nevertheless the electrons strongly affect the nonlinear dynamics, due to their damping which relaxes in time when their profile radially redistributes.

In summary, the contribution of the electron redistribution to the nonlinear saturation of BAEs and consequent excitation of ZFs is proven here to be important in explaining the higher saturation levels of BAEs and ZFs observed in fully nonlinear simulations, with respect to simulations where only the wave-particle nonlinearity with the EPs is considered.

As future steps, the importance of the electron damping in fully nonlinear simulations will be studied in more experimentally relevant magnetic configurations, and compared with the experimental measurements. Moreover, the inclusion of the interaction with higher-n modes like ITG modes will be described.

\section*{Acknowledgments}

%\vskip 1em

Interesting discussions with F. Zonca, Z. Qiu, L. Villard, F. Jenko, I. Novikau and A. Di Siena are gratefully acknowledged. This work has been carried out within the framework of the EUROfusion Consortium and has received funding from the Euratom research and training program 2014-2018 and 2019-2020 under grant agreement No 633053 within the framework of the {\emph{Multiscale Energetic particle Transport}} (MET) European Eurofusion Project. The views and opinions expressed herein do not necessarily reflect those of the European Commission. Simulations were performed on the HPC-Marconi supercomputer within the framework of the ORBFAST and OrbZONE projects.

%\newpage
\appendix
\addcontentsline{toc}{chapter}{APPENDICES}

\section{Convergence with the electron mass}
\label{app:mass}

\begin{figure}[h!]
\begin{center}
%\vskip -1em
\includegraphics[width=0.65\textwidth]{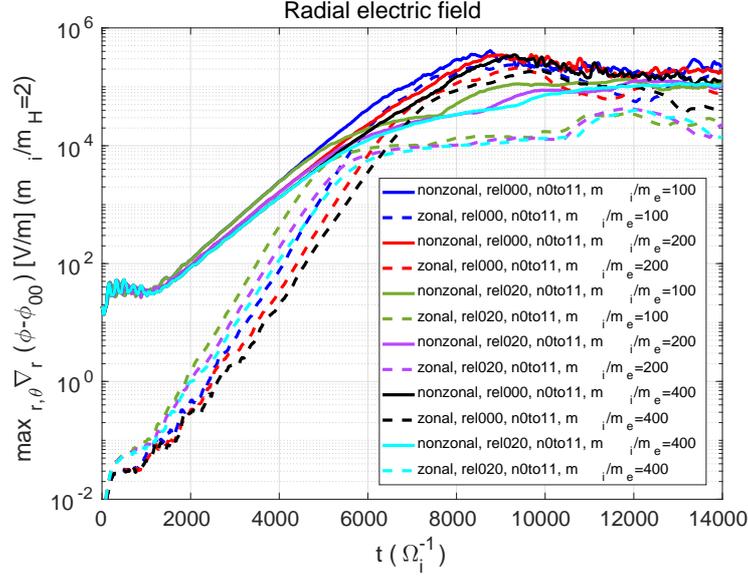}
%\vskip -1em
\caption{\label{fig:mass-scan} Nonzonal (continuous lines) and zonal (dashed lines) radial electric field. rel000 denotes fully nonlinear sims. rel020 denotes sims where the electrons are pushed along unperturbed trajectories. Note that the convergence on the mass ratio is achieved.}
\end{center} 
\vskip -1em
\end{figure}

Here we show the convergence scan with the mass ratio, for sims with $\langle n_{EP}\rangle /\langle n_e\rangle =0.03$. In Fig.~\ref{fig:mass-scan} we can see that going from $m_i/m_e=100$ to $m_i/m_e=200$ we still see some slight difference in the linear growth rates, which becomes even smaller from $m_i/m_e=200$ to $m_i/m_e=400$. The saturation levels, on the other hand, seem unchanged. Therefore, the converence is achieved.

\end{document}